\documentclass[12pt]{article}
\usepackage{amsmath, mathtools, color}
\usepackage{float}

\usepackage[round,authoryear,comma,sort&compress,numbers]{natbib}       

\usepackage{accents}
\usepackage{stackengine}

\usepackage{graphicx, geometry,graphicx}
\usepackage{epstopdf}
        \usepackage{dutchcal}

\newcommand{\be}{\begin{equation}}
\newcommand{\en}{\end{equation}}

\renewcommand{\vec}[1]{\boldsymbol{#1}}

\def \curl{\mbox{curl\hskip 1pt}}

\def \div{\mbox{div\hskip 1pt}}

\def \grad{\mbox{grad\hskip 1pt}}

\begin{document}

\title{Pattern evolution\\ in bending dielectric-elastomeric bilayers\footnote{Dedicated in friendship and esteem to Davide Bigoni, a master of stability analysis}}

\author{Yipin Su$^{1,2}$, Bin Wu$^3$, Weiqiu Chen$^1$, Michel Destrade$^{2,1}$
\\[12pt]
$^1$Department of Engineering Mechanics,\\
Zhejiang University, Hangzhou 310027, P.R. China;\\[12pt]
$^2$School of Mathematics, Statistics and Applied Mathematics, \\
NUI Galway, University Road, Galway, Ireland;\\[12pt]
$^3$Department of Mechanical and Aerospace Engineering,\\
Politecnico di Torino, Torino 10129, Italy.}

\date{}
%
\maketitle

\begin{abstract}
We propose theoretical and numerical {\color{black}analyses} of smart bending deformation of a dielectric-elastic  bilayer in response to a voltage, based on the nonlinear theory of electro-elasticity and the associated linear{\color{black}ized} incremental field theory.
We reveal that the mechanism allowing the bending angle of the bilayer can be tuned by adjusting the applied voltage.
Further{\color{black}more}, we investigate how much can the bilayer be bent before it loses its stability by buckling when one of its faces is under too much compression. 
We find that the physical properties of the two layers must be selected to be of the same order of magnitude to obtain a consequent bending without encountering buckling. 
If required, the wrinkles can be designed to appear on either the inner or the outer bent surface of the buckled bilayer. 
We validate the results through comparison with those of the classical elastic problem.
\end{abstract}


\newpage


\section{Introduction}

\begin{figure}[h!]
\centering
\includegraphics[width=0.8\textwidth]{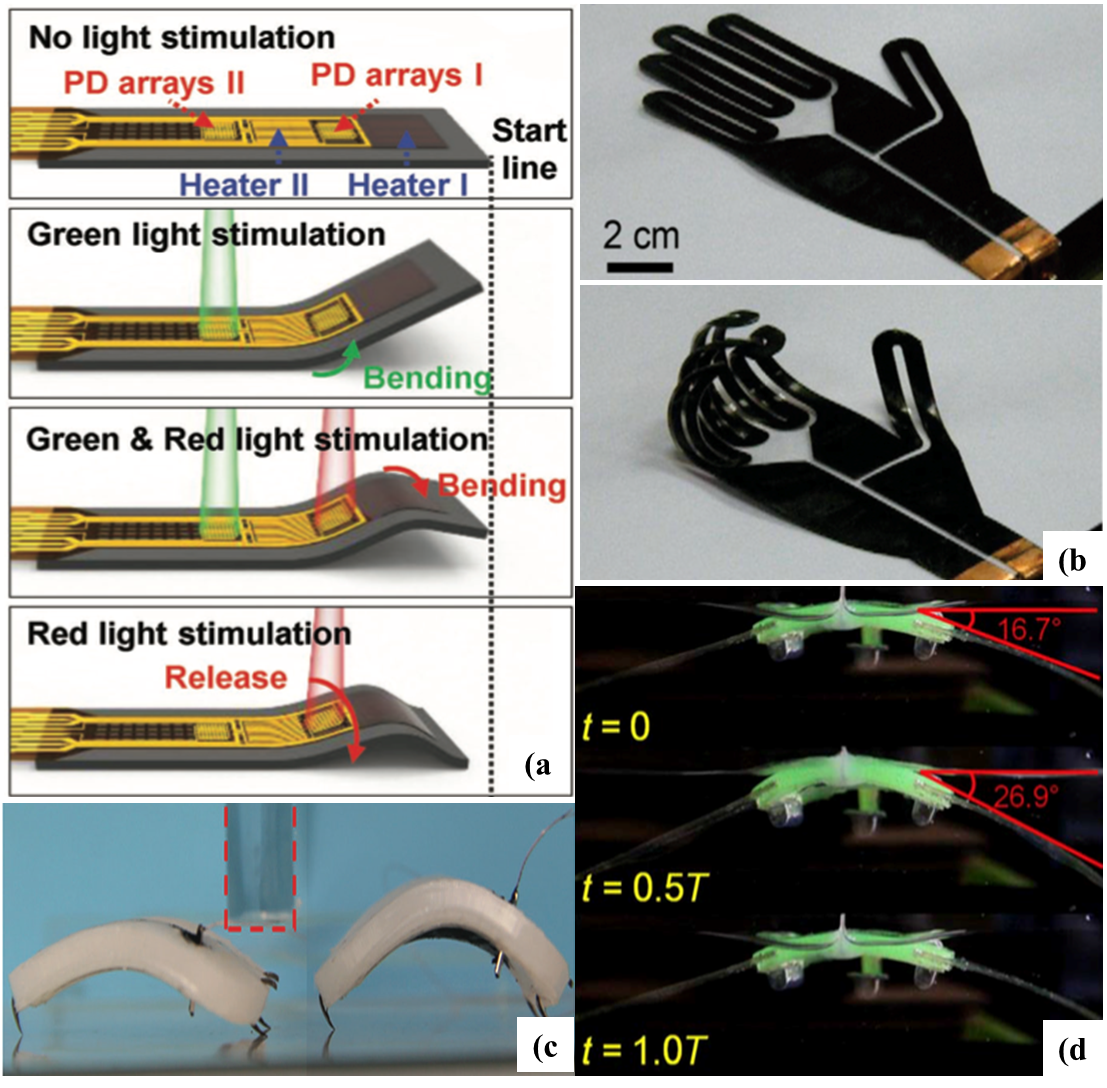}
\caption{
{\footnotesize
Smart bending deformations in intelligent devices: (a) A soft robot mimicking the crawling motion of an inchworm: bending deformations occur in response to the alternating green and red laser lights and result in locomotion of the bimorph \citep{Wang2018}; (b) Bending deformation of a hand-shaped actuator, where the bending angle of each finger can be tuned by the applied voltage \citep{Li2015}; (c) A dielectric elastomer actuated soft robot can sneak under a barrier (red dashed line): here the wormlike self-walking motion is induced by tuning the voltage-induced bending angle of the elastomer \citep{Li2018}; (d) A self-powered soft robot made of dielectric elastomer and ionically conductive hydrogel mimicks a swimming fish by periodically bending its fins \citep{Li2017} .
}
}
\label{figure1}
\end{figure}

Finite bending is a most common deformation of solid{\color{black}s} in nature and in engineering applications, generally triggered by the application of mechanical loads.
We call this scenario, \emph{mechanical bending}. 
In contrast, we use the denomination \emph{smart bending} to refer to the bending deformation of solid{\color{black}s} that is intentionally activated by other controllable stimuli such as pH, temperature, electric field, magnetic field or light signals. 
Smart bending provides an effective way to realize intelligent control and has potential applications in actuators, soft robots, flexible electronics, bionic engineering, etc.  
Various controllable motions such as large-strain bending, helical curling, climbing, and even bionic actuations (wormlike self-walking motion and imitating human-hand actions) have been brought {\color{black}in}to life recently by   smart bending, see examples in Figure \ref{figure1}. On the other hand, little attention has been devoted to the theoretical investigation of the smart bending deformation of solids, and the physical mechanisms at play to achieve large and stable smart bending deformations of intelligent structures are still not completely understood.

A clever strategy to realize smart bending of a solid is to rely on a layered structure which generates bending moments due to the spatially \color{black}differential \color{black} stresses along its thickness {\color{black}induced by} external stimuli \citep{Morimoto2015, Wang2017, Nardinocchi2017}. 
Motivated by the explosion of applications of dielectric elastomers --  smart soft materials capable of  large actuation deformation in fast response to electrical simulation -- in actuators, soft robots and bioengineering \citep{Pelrine2000, O'Halloran2008, Brochu2010}, the aim of this work is to design a dielectric-elastomeric bilayer  able to undergo large voltage-induced smart bending, and to reveal the physical mechanisms enabling a considerable finite bending deformation of the structure.

In this paper, we first propose a theoretical analysis of finite smart bending deformation of a dielectric-elastic bilayer in response to an applied voltage.
\color{black}
The main idea is that by controlling the applied voltage, we control the extend of the bending deformation and thus achieve (theoretically) our goal of `smart bending'.
\color{black}
Here we adopt a nonlinear field theory \citep{Dorf05, Dorf06, Suo08} to describe the finite deformation coupled with an electric field, and find the optimal parameters for the bilayer to achieve a large bending angle.

Next, it is well known that \color{black} theoretically, \color{black} bending instability may occur in an extremely bent elastic solid \citep{Haughton99, Destrade09}. 
On the one hand, instability has long been recognized as a failure mode of structures and thus should be carefully analysed \citep{Madsen2006}. 
The onset of  {\color{black}bending} instability can be complicated by a layered structure, where many different routes can be taken to reach wrinkling, as shown by Bigoni and co-workers \citep{Rocca10,Rocca11,Bigoni12}.
On the other hand, it was recently identified  that wrinkles could be beneficial and harvested to generate complex 3D architectures \citep{Wu2013, Liu2016, Fu2018}. 

In this paper, we use the linearized incremental theory \citep{Dorf10, Bertoldi2011, Rudykh2017, Shmuel2018, Su2018b} to further study the possible wrinkling of the bilayer caused by the large bending deformation.

The paper is structured as follows. 
In Section \ref{section2}, we adopt the multiplicative decomposition method to derive the governing equations of the voltage-controlled bending deformation of dielectric-elastomeric bilayers. 
We then conduct a linearized stability analysis in Section \ref{section3} and use the surface impedance matrix method to predict the onset of wrinkles on the bent faces of the bilayer.  
The formulations developed in Section \ref{section2} and Section \ref{section3} are valid for bilayers modeled by any form of energy function. 
In Section \ref{section4} we present numerical calculations for bilayers modelled with the Gent strain energy function (a model often used for strain-stiffening solids) to study the voltage-controlled bending deformation and its associated wrinkling pattern evolution. 
We find that the physical parameters of the dielectric and elastic elastomers must be selected to be of  the same order of magnitude to obtain a considerable bending angle.
We also find that the wrinkles of an extremely bent bilayer can be intentionally designed to occur on either the inner or the outer bent face.
Finally we draw some conclusions in Section \ref{section5}.


\section{Voltage-induced finite bending of a dielectric-elastomeric  {\color{black}bilayer}}
\label{section2}


\begin{figure}[h!]
\centering
\includegraphics[width=0.8\textwidth]{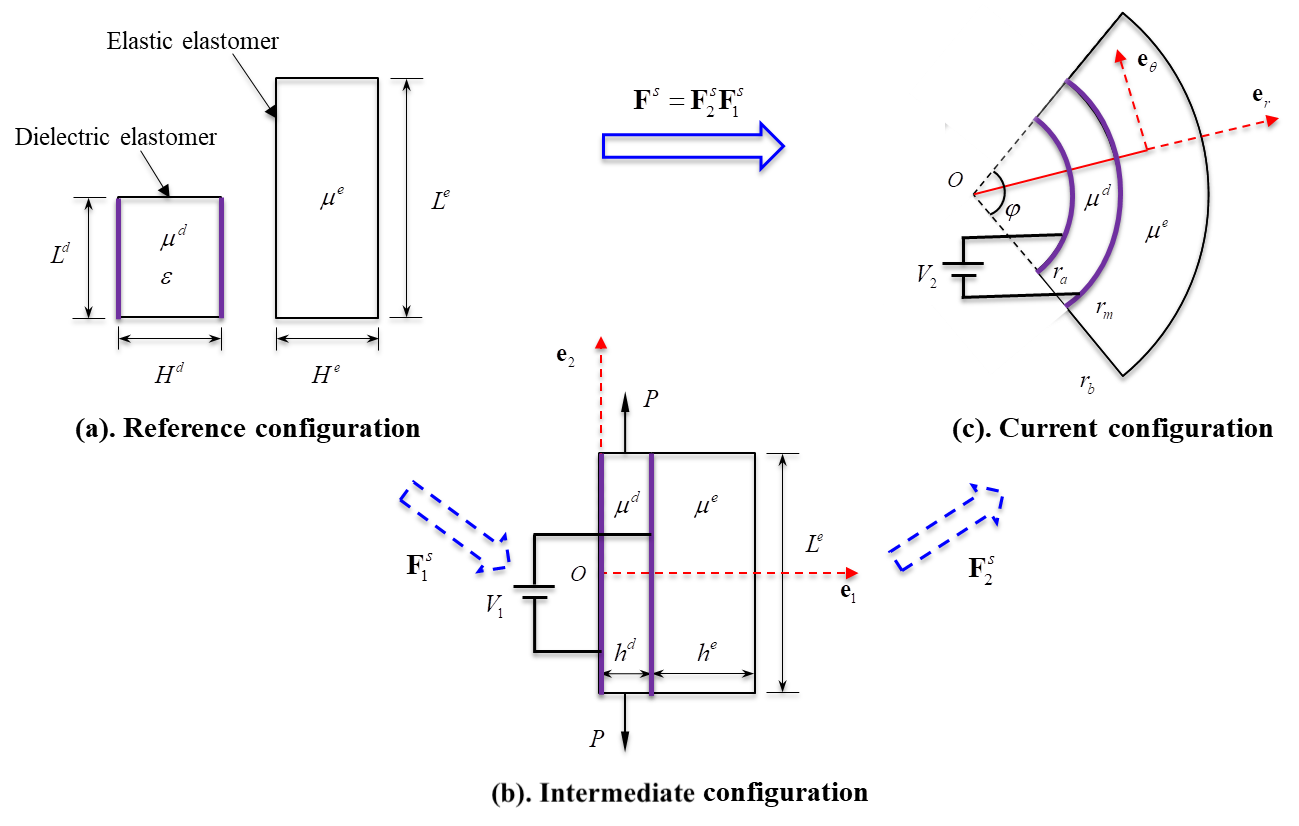}
\caption{
{\footnotesize
Voltage-controlled bending deformation of a dielectric-elastic bilayer induced by inhomogeneous deformation and its multiplicative decomposition.
}
}
\label{figure2}
\end{figure}

Figure \ref{figure2} illustrates the mechanism at play in  the dielectric-elastomeric bilayer and its bending. 
Consider two initially undeformed incompressible elastomers, one dielectric and the other elastic, whose physical parameters are indicated henceforth with the superscripts `$d$' and `$e$', respectively. 
The initial thickness and length of the $s$-th ($s=d,e$) block are $H^{s}$ and $L^{s}$, respectively, $\mu^{s}$  {\color{black}denotes} its initial mechanical shear modulus, and $\varepsilon$ is the permittivity of the dielectric block.

Subject to a voltage $V_1$ along the thickness and a pre-stress $P$ along the length, the dielectric block can be stretched to have the same length as the undeformed elastic block. 
Then the dielectric block is assumed to be perfectly bonded  to the elastic block along their length to form a bilayer (i.e., the  stresses and displacements of the two elastomers are continuous at the interface). 
By releasing the pre-stress and reducing\color{black}, in a controlled manner, \color{black} the voltage to $V_2$ ($<V_1$), an inhomogeneous deformation takes place in the bilayer, which can result in a global bending, with bending angle $\varphi$, say.

\color{black}
In principle, we could induce the same mismatch by  applying a pure mechanical pre-stress on the dielectric block, and a bending deformation of the bilayer would then result from releasing that stress. 
However, that ‘pure mechanical bending’ deformation, in contrast to our ‘smart bending’ deformation, is difficult to control in practice. 
Here we show that the voltage-induced bending deformation of the bilayer can be tuned effectively by varying the applied voltage only, which provides an effective way to realize intelligent control. Note that when the bilayer bends, the applied voltage is not completely off. Instead, it is reduced (eventually all the way to zero if desired) in a controlled way.
We are thus solving an electromechanical coupling nonlinear problem, which has not been analyzed in literature before as far as we know.
\color{black}

The in-plane width of each block $A^d=A^e=A$ is taken to be much longer than its thickness and length, and so we make the \emph{plane strain} deformation assumption to  simplify the mathematical modeling \citep{Morimoto2015, Wang2017}. 
We also assume that the materials are \emph{incompressible}.

The total deformation gradient $\vec F^{s}$ of the bilayer, from the initial undeformed state to the current state of bending, is decomposed into two parts as shown in Figure \ref{figure2}: $\vec F^{s}$=$\vec F^{s}_2 \vec F^{s}_1$, where $\vec F^{s}_1$ is the (homogeneous) deformation gradient from the reference  to the intermediate configuration, and $\vec F^{s}_2$ is the (inhomogeneous) deformation gradient of the deformation from the intermediate to the current configuration.

We first consider the homogeneous deformation of the layers from the reference configuration to the intermediate configuration, with the deformation gradients 
\begin{equation}\label{deformation1}
\vec F^{e}_1=\vec I, \qquad
\vec F^{d}_1=\mathrm {diag}\left [(\lambda^{d}_1)^{-1},\lambda^{d}_1 ,1\right], 
\end{equation} 
due to the plane strain assumption and the incompressibility of the materials.
The deformed thicknesses and lengths are $h^e = H^e$,  $h^d=(\lambda^d_1)^{-1}H^d$ and $l^e = l^d =L^e = \lambda^d_1 L^d$, respectively. 

Here, the stretch along the length of the dielectric block $\lambda^d_1=\lambda^d_1(V_1, P)$, subject to the combined action of voltage $V_1$ and pre-stress $P$, can be determined by solving the equation
\begin{equation}\label{voltage-displacement1}
\frac{\partial W^d_1}{\partial \lambda^d_1}(\lambda^d_1,D_1)=\color{black}P\color{black},
\end{equation} 
where $W^d_1(\lambda^d_1,D_1)$ is the energy density of the dielectric elastomer and $D_1$ is the only non-zero component of the Lagrangian electric displacement along the thickness.
It is connected with the applied voltage $V_1$ through
\begin{equation}\label{voltage1}
\frac{V_1}{H^d}=\frac{\partial W^d_1}{\partial D_1}(\lambda^d_1,D_1).
\end{equation} 

Following the homogeneous deformation $\vec F^s_1$, the bilayer occupies the region
\begin{equation}
0 \le X_1 \le h^d+h^e,
\quad
-L^e/2 \le X_2 \le L^e/2,
\quad
-A/2 \le X_3 \le A/2,
\end{equation}
as depicted in Figure \ref{figure2}. 

We then consider the second (inhomogeneous) deformation stage from the intermediate to the current configuration. After releasing $P$ and reducing $V_1$, the bilayer bends and occupies the current region
\begin{equation}
r_a \le r \le r_b,
\quad
-\varphi/2 \le \theta \le \varphi/2,
\quad
-A/2 \le z \le A/2,
\end{equation}
by undergoing the following bending deformation \color{black} \citep{Rivlin1949, Green54, Ogden97}\color{black}
\begin{equation} \label{bending}
r=\sqrt{D+2X_1/\omega}, \quad \theta=\omega X_2, \quad z=X_3.
\end{equation} 
Here $(X_1,X_2,X_3)$ and $(r,\theta, z)$ are the rectangular Cartesian and cylindrical coordinates giving the position of an arbitrary  point of the solid in the intermediate and current configurations, with orthogonal bases $(\vec e_1,\vec e_2,\vec e_3)$ and $(\vec e_r,\vec e_\theta,\vec e_z)$, respectively, and $D$ and $\omega$ are constants to be determined. 
The inner, interfacial and outer radii and the bending angle of the deformed sector $r_a, r_m, r_b$ and $\varphi$ are given by 
\begin{equation}\label{parameters}
r_a=\sqrt{D},
\qquad
r_m=\sqrt{D+\frac{2h^d}{\omega}},
\qquad
r_b=\sqrt{D+\frac{2(h^d+h^e)}{\omega}},
\quad
\varphi=\omega L^e,
\end{equation}
respectively. \color{black}Note that the continuity of the displacement field is implicit in Eq. \eqref{parameters}. \color{black}
Then the deformation gradient of the bilayer in this stage is
\begin{equation}\label{deformation2}
\vec F_2^s=\mathrm{diag}\left[(\lambda^s_2)^{-1},\lambda^s_2,1\right],
\end{equation}
in the $\vec e_i \otimes \vec e_j$ \color{black}($i=r,\theta, z$, $j=1,2,3$) \color{black} basis, where $\lambda^s_2=\omega r$ is the circumferential stretch of the solid in the current configuration.

Consequently, from Eqs. \eqref{deformation1} and \eqref{deformation2} the total deformation gradient of the bilayer from the reference configuration to the current configuration reads 
\begin{equation}\label{total-deformation}
\vec F^{s}=\vec F^{s}_2 \vec F^{s}_1=\mathrm {diag}\left[(\lambda^{s})^{-1},\lambda^{s} ,1\right],
\end{equation}
where $\lambda^s=\lambda^s_1\omega r$ is the total circumferential stretch of the $s$-th block; here $\lambda^d=\lambda^d_1\omega r$ and $\lambda^e=\omega r$.

From \color{black} Eqs.\eqref{parameters}-\eqref{total-deformation} \color{black} we derive  the  relationships
\begin{equation}\label{interfacial2}
\lambda^d_m=\lambda^d_1\lambda^e_m, 
\qquad 
\frac{\left(\lambda^e_m\right)^2-\left(\lambda_a/\lambda^d_1\right)^2}{\left(\lambda_b\right)^2-\left(\lambda_a/\lambda^d_1\right)^2}=\frac{1}{1+\lambda^d_1\overline H},
\end{equation}
where $\overline H=H^e/H^d$ is the ratio of the thicknesses of the elastic and dielectric elastomers, $\lambda_a=\lambda^d_1\omega r_a$, $\lambda_b=\omega r_b$ are the inner and outer circumferential stretches of the solid, and $\lambda_m^d=\lambda^d_1\omega r_m$, $\lambda_m^e=\omega r_m$ are the circumferential stretches of the dielectric and elastic elastomers at the interface, respectively. 
Note that the circumferential stretches of the two elastomers at the interface, $\lambda^d_m$ and $\lambda^e_m$, are not the same.

According the theory of nonlinear electro-elasticity, the stresses in the radial and hoop directions of the solid are obtained as \citep{Su2018}
\begin{align}\label{normal-stress}
& \tau^d_{rr}=W^d(\lambda^d,D_2)-W^d(\lambda_a,D_2), &
& \tau^d_{\theta\theta}=\lambda^d\frac{\partial W^d}{\partial \lambda^d} (\lambda^d,D_2) + W^d(\lambda^d,D_2)-W^d(\lambda_a,D_2), \\ \notag
& \tau^e_{rr}=W^e(\lambda^e)-W^e(\lambda_b), &
& \tau^e_{\theta\theta}=\lambda^e\frac{\partial W^e}{\partial \lambda^e}(\lambda^e) + W^e(\lambda^e)-W^e(\lambda_b), 
\end{align}
where $W^d(\lambda^d,D_2)$ and $W^e(\lambda^e)$ are the  energy densities of the  dielectric and elastic elastomers, respectively, and $D_2$ is the only non-zero component of the Lagrangian electric displacement along the thickness associated with the applied voltage $V_2$. 
\color{black}Note that we have used the equilibrium equation in the radial direction\color{black}, and the traction-free conditions $\tau_{rr}^d(r_a)=\tau_{rr}^e(r_b)=0$ at the inner and outer surfaces $r =r_a, r_b$ of the bilayer to derive Eq. \eqref{normal-stress}.

The continuity of the radial stress field at the bonded interface between the dielectric and elastic blocks reads
\begin{equation}\label{interfacial-stretch}
\tau_{rr}^d(r_m)=\tau_{rr}^e(r_m).
\end{equation} 

Since there is no mechanical moment applied on the lateral boundaries $\theta=\pm \varphi/2$ of the bilayer, we have
\begin{equation}\label{interfacial-moment}
\int_{r_a}^{r_m}\tau_{\theta\theta}^d r\,\text dr+\int_{r_m}^{r_b}\tau_{\theta\theta}^e r\,\text dr=0.
\end{equation} 
Further, we showed elsewhere \citep{Su2018} that no normal force is required on the end faces to {\color{black}{effect}} the bending, so that 
\begin{equation}
\int_{r_a}^{r_m}\tau_{\theta\theta}^d\,\text dr+\int_{r_m}^{r_b}\tau_{\theta\theta}^e\,\text dr=0.
\end{equation} 
Finally, the voltage difference between the two faces of the deformed dielectric elastomer is obtained as \citep{Wu2017, Su2018}
\begin{equation}\label{connection}
V_2=\int_{r_a}^{r_m}E_r\,\text dr=\int_{r_a}^{r_m}\lambda^d\frac{\partial W^d}{\partial D_2}\,\text dr,
\end{equation}
where $E_r=\lambda^d \partial W^d/\partial D_2$ is the true radial electric field in the deformed dielectric layer.

Summing up, we can determine $\lambda_1^d, \lambda_a, \lambda^d_m, \lambda^e_m$ and $\lambda_b$ by solving \color{black}simultaneous \color{black} Eqs. \eqref{voltage-displacement1}, \eqref{voltage1}, \eqref{interfacial2}, \eqref{normal-stress}, \eqref{interfacial-stretch}, \eqref{interfacial-moment} and \eqref{connection}, once the loadings $P$, $V_1$ and $V_2$ are prescribed. 
We then obtain the bending angle  from Eq. \eqref{parameters}$_4$  and the stress distribution in the bilayer from Eq. \eqref{normal-stress}.


\section{Linearized stability analysis}
\label{section3}


We adopt the linearized incremental theory \citep{Dorf10, Bertoldi2011, Rudykh2017, Shmuel2018, Su2018b} to model the onset of buckling of the deformed bilayer and focus on the case when the applied voltage is completely removed ($V_2=0$). 
Readers may refer to the paper by \cite{Su2018} for a detailed derivation of the formulas for a single dielectric elastomer.
Here we just give the general equations governing the deformation of a small-amplitude perturbation superposed on the deformed dielectric elastomer; the incremental  equations for the elastic elastomer are found from those for the dielectric elastomer by making the electric field vanish. 
Thus we omit the superscript in the following for ease of reading; and we will revert to the superscript notation to describe the boundary conditions and the interfacial condition{\color{black}{s}}.


\subsection{Incremental deformation}


Upon the deformed configuration, we superimpose a small incremental displacement $\vec u$ and a small incremental (true) electric displacement $\dot{\vec D}$. 
Here and in the following the incremental quantities are denoted by a dot. 

The incremental motion of the dielectric block is governed by the incremental equations of equilibrium and of incompressibility, which read
\begin{equation}\label{balance}
\div \vec{\dot T}=\vec 0, \quad \div \vec{\dot D}=0, \quad \curl \vec{\dot E}=0, \quad \div \vec u=0.
\end{equation}
Here $\vec{\dot T}= \boldsymbol{\mathcal A}(\grad \vec u) +\vec \Gamma\vec{\dot D}+p(\grad \vec u)-\dot{p}\vec I$ and $\vec{\dot E}=\vec \Gamma^T(\grad \vec u)+\vec K\vec{\dot D}$ are the incremental forms of Cauchy stress and true electric field, respectively, where $p$ is a Lagrange multiplier associated with the incompressibility constraint and the fourth-, third- and second-order tensors $\boldsymbol{\mathcal A}$, $\vec \Gamma$ and $\vec K$ are the effective electro-elastic moduli. 

According to Eq.\eqref{balance}$_3$, we may introduce an incremental electric potential $\dot \phi$ to rewrite the incremental electric field as $\dot{\vec E}=-\grad\dot{\phi}$.

We assume that there is no incremental traction on the inner and outer surfaces of the bent bilayer. 
The applied voltage is taken to be a constant to yield the following boundary conditions
\begin{align}\label{incrementalbc}
\dot{T}^d_{rr}(r_a)=\dot{T}^d_{r\theta}(r_a)=\dot{\phi}^d(r_a)=\dot{T}^e_{rr}(r_b)=\dot{T}^e_{r\theta}(r_b)=\dot{\phi}^d(r_m)=0.
\end{align} 
Finally, because the two blocks are perfectly bonded, we have 
\begin{equation}\label{incremental-interface}
\dot{T}^d_{rr}(r_m)=\dot{T}^e_{rr}(r_m), \quad \dot{T}^d_{r\theta}(r_m)=\dot{T}^e_{r\theta}(r_m).
\end{equation}


\subsection{The surface impedance method}


The solution for the linear equations of the incremental deformation may be assumed in the form \citep{Su2016}
\begin{align}\label{incremental-solutions}
&u_r=U_r(r)\text{cos}\left(n\theta\right), & \quad & u_\theta=U_\theta(r)\text{sin}\left(n\theta\right), & \quad & \dot{\phi}=\Phi(r)\text{cos}\left(n\theta\right), \notag \\& \dot T_{rr}=\Sigma_{rr}(r)\text{cos}\left(n\theta\right), & \quad & \dot T_{r\theta}=\Sigma_{r\theta}(r)\text{sin}\left(n\theta\right),
& \quad & \dot D_{r}=\Delta_r(r)\text{cos}\left(n\theta\right),
\end{align}
where  $U_r, U_\theta, \Phi, \Sigma_{rr}, \Sigma_{r\theta}$ and $\Delta_r$ are scalar functions of $r$ only, and $n$ is the circumferential wave number.
It can be determined from the incremental traction free boundary conditions on the end faces $\theta=\pm \varphi/2$ as
$n=2q\pi/\varphi$ \citep{Destrade09}, where the positive integer $q$ gives the number of circumferential wrinkles of the solid when buckling occurs. 
 
Then the governing equations of the incremental motion can be recast in the form of a  first-order differential system, as
\begin{equation}\label{stroh}
\frac{\text{d}}{\text{d}r}\vec \eta(r)=\frac{1}{r}\vec G\vec\eta(r)=\frac{1}{r}
\left[ \begin{matrix}
 \vec G_1 & \vec G_2 \\
 \vec G_3 & \vec G_4\end{matrix} \right]
\vec\eta(r),
\end{equation}
where
$\vec\eta(r)=\left[ \begin{matrix}
 U_r & U_\theta &  r\Delta_r & r\Sigma_{rr} &r\Sigma_{r\theta} & \Phi
\end{matrix} \right]^{\text T}=\left[ \begin{matrix}\vec U & \vec S\end{matrix} \right]^{\text T}$
is the Stroh vector (with $\vec U=\left[ \begin{matrix}
 U_r & U_\theta & r\Delta_r\end{matrix} \right]^{\text T}$ and $\vec S=\left[ \begin{matrix}
r\Sigma_{rr} &r\Sigma_{r\theta} & \Phi 
\end{matrix} \right]^{\text T}$), $\vec G$ is the so-called Stroh matrix, with four $3\times 3$ sub-blocks $\vec G_1$, $\vec G_2$, $\vec G_3$ and $\vec G_4$, given by
\begin{align}
&\vec G_1=\left[ \begin{matrix}
 -1 & -n & 0\\
 \frac{n\left(\gamma_{12}-\tau_{rr}\right)}{\gamma_{12}} &  \frac{\gamma_{12}-\tau_{rr}}{\gamma_{12}}  & 0\\
-\frac{n^2 \tau_{rr}}{\gamma_{12}}\frac{\Gamma_{122}}{K_{22}} & -\frac{n\tau_{rr}}{\gamma_{12}}\frac{\Gamma_{122}}{K_{22}} & 0
\end{matrix} \right], \quad
\vec G_2=\left[\begin{matrix}
0 & 0 & 0 \\
0 & \frac{1}{\gamma_{12}} & -\frac{n}{\gamma_{12}}\frac{\Gamma_{122}}{K_{22}} \\
0 & \frac{n}{\gamma_{12}}\frac{\Gamma_{122}}{K_{022}}& -\frac{n^2}{K_{22}}\left(1+\frac{\Gamma_{122}^2}{K_{22}}\frac{1}{\gamma_{12}}\right)
\end{matrix}\right], \notag \\[6pt]
&\vec G_3=\left[\begin{matrix}
\kappa_{11} & \kappa_{12} &-\left(\Gamma_{111}-\Gamma_{221}\right) \\
\kappa_{12} & \kappa_{22} &-n\left(\Gamma_{111}-\Gamma_{221}\right) \\
\Gamma_{111}-\Gamma_{221} & n\left(\Gamma_{111}-\Gamma_{221}\right) & -K_{11}
\end{matrix}\right], \notag \\[6pt]
&\vec G_4=\left[ \begin{matrix}
 1 & -\frac{n\left(\gamma_{12}-\tau_{rr}\right)}{\gamma_{12}} & -\frac{n^2 \tau_{rr}}{\gamma_{12}}\frac{\Gamma_{122}}{K_{22}}\\
 n &  -\frac{\gamma_{12}-\tau_{rr}}{\gamma_{12}} & -\frac{n\tau_{rr}}{\gamma_{12}}\frac{\Gamma_{122}}{K_{22}}\\
 0 & 0 & 0
\end{matrix} \right].
\end{align}
Here{\color{black}{, we have}}
\begin{align}\label{material-parameters}
&\gamma_{12}= \mathcal A_{1212}-\frac{\Gamma_{122}^2}{K_{22}}, \quad \gamma_{21}= \mathcal A_{2121}-\frac{\Gamma_{122}^2}{K_{22}}, \notag \\
& \beta_{12}=\frac{1}{2}\left(\mathcal A_{1111}+\mathcal A_{2222}-2\mathcal A_{1122} - 2 \mathcal A_{1221} + \frac{2\Gamma_{122}^2}{K_{22}}\right), \notag \\
&\kappa_{11}=2(\gamma_{12}-\tau_{rr}+\beta_{12})+n^2\left[\gamma_{21}-\frac{\left(\gamma_{12}-\tau_{rr}\right)^2}{\gamma_{12}}\right], \notag \\
&\kappa_{12}=n\left(\gamma_{12}+\gamma_{21}+2\beta_{12}-\frac{\tau_{rr}^2}{\gamma_{12}}\right),\notag\\ 
&\kappa_{22}=2n^2(\gamma_{12}-\tau_{rr}+\beta_{12})+\gamma_{21}-\frac{\left(\gamma_{12}-\tau_{rr}\right)^2}{\gamma_{12}}.
\end{align}

Note that by setting the voltage appearing in the Stroh matrix for the dielectric elastomer to be zero, we  obtain the Stroh matrix for the purely elastic elastomer.

Now the incremental boundary conditions \eqref{incrementalbc} read
\begin{equation}\label{boundary-Stroh}
\vec S^d(r_a)=\vec S^e(r_b)=\vec 0,
\end{equation}
and the incremental interfacial conditions \eqref{incremental-interface} read
\begin{equation}\label{boundary-interfacial}
\vec S^d(r_m)=\vec S^e(r_m), \quad \vec U^d(r_m)=\vec U^e(r_m).
\end{equation}

Define a conditional impedance matrix $\vec z^s(r,r_a)$ such that
\begin{equation}\label{stress-Stroh}
\vec S^s(r)=\vec z^s(r,r_a)\vec U^s(r),
\end{equation}
Then we use the surface impedance matrix method \citep{Destrade09, Du2018} to rewrite the Stroh differential system \eqref{stroh} as
\begin{equation}\label{two_Riccati}
\frac{\text d}{\text dr}\vec U^s=\frac{1}{r}\vec G^s_1 \vec U^s+\frac{1}{r}\vec G^s_2 \vec z^s \vec U^s, \quad
\frac{\text d}{\text dr}(\vec z^s \vec U^s)=\frac{1}{r}\vec G^s_3 \vec U^s+\frac{1}{r}\vec G^s_4 \vec z^s \vec U^s.
\end{equation}
The following Riccati differential equation can be obtained by elimination of $\vec U^s$ from Eq. \eqref{two_Riccati}
\begin{equation}\label{Riccati}
\frac{\text d\vec z^s}{\text dr}=\frac{1}{r}\left(-\vec z^s \vec G^s_1-\vec z^s \vec G^s_2 \vec z^s+\vec G^s_3+\vec G^s_4 \vec z^s\right).
\end{equation}

First we integrate Eq. \eqref{Riccati} numerically for the dielectric layer, from $r_a$ to $r_m$ with the initial condition $\vec z^d(r_a,r_a)=\vec 0$,  to obtain $\vec z^d(r_m,r_a)$. 
Then we integrate Eq. \eqref{Riccati} numerically for the elastic layer, from $r_m$ to $r_b$  with the initial condition $\vec z^e(r_m,r_a)=\vec z^d(r_m,r_a)$,  to obtain $\vec z^e(r_b,r_a)$. 
Finally we tune the applied voltage $V_1$ until the following target condition is satisfied
\begin{equation}\label{dispersion}
\text{det }\vec z^e(r_b,r_a)=0,
\end{equation}
which results from the boundary condition $\vec S^e(r_b)=\vec 0$.

The conclusion is that for a dielectric-elastic bilayer with prescribed physical properties, the critical applied voltage $V_{1c}$ of instability can be determined numerically, and so can the critical values of the bending angle $\varphi_c$ and of the stretches $\lambda_{1c}^d, \lambda_{ac}, \lambda^d_{mc}, \lambda^e_{mc}$ and $\lambda_{bc}$.

It follows from the condition $\vec S^e(r_b)=\vec z^e(r_b,r_a)\vec U^e(r_b)=\vec 0$ that the ratios of the incremental motion on the outer face of the sector $t_\theta=U_\theta(r_b)/U_r(r_b)$, $t_\Phi=\Phi(r_b)/U_r(r_b)$ can be determined once $\vec z^e(r_b,r_a)$ is known. Now that  the critical voltage $V_{1c}$ has been obtained, we can define the other conditional impedance matrix $\vec z^s(r,r_b)$ in a similar way and integrate simultaneously the corresponding Eqs. \eqref{two_Riccati} and \eqref{Riccati} from $r_b$ to $r_a$ with the following initial conditions
\begin{equation}
\vec U(r_b)=U(r_b)\left[\begin{matrix}1 & t_\theta & t_\Phi \end{matrix}\right]^{\text T}, \quad \vec z^e(r_b, r_b)=\vec 0,
\end{equation}
where $U(r_b)$ is the arbitrary amplitude of the radial displacement on the outer surface.
This integration determines the full distribution of the incremental field $\vec U^s(r)$ in the deformed sector and the corresponding buckling pattern.


\section{Numerical results}
\label{section4}



\subsection{Material model}

\begin{figure}[h!]
\centering
\includegraphics[width=0.7\textwidth]{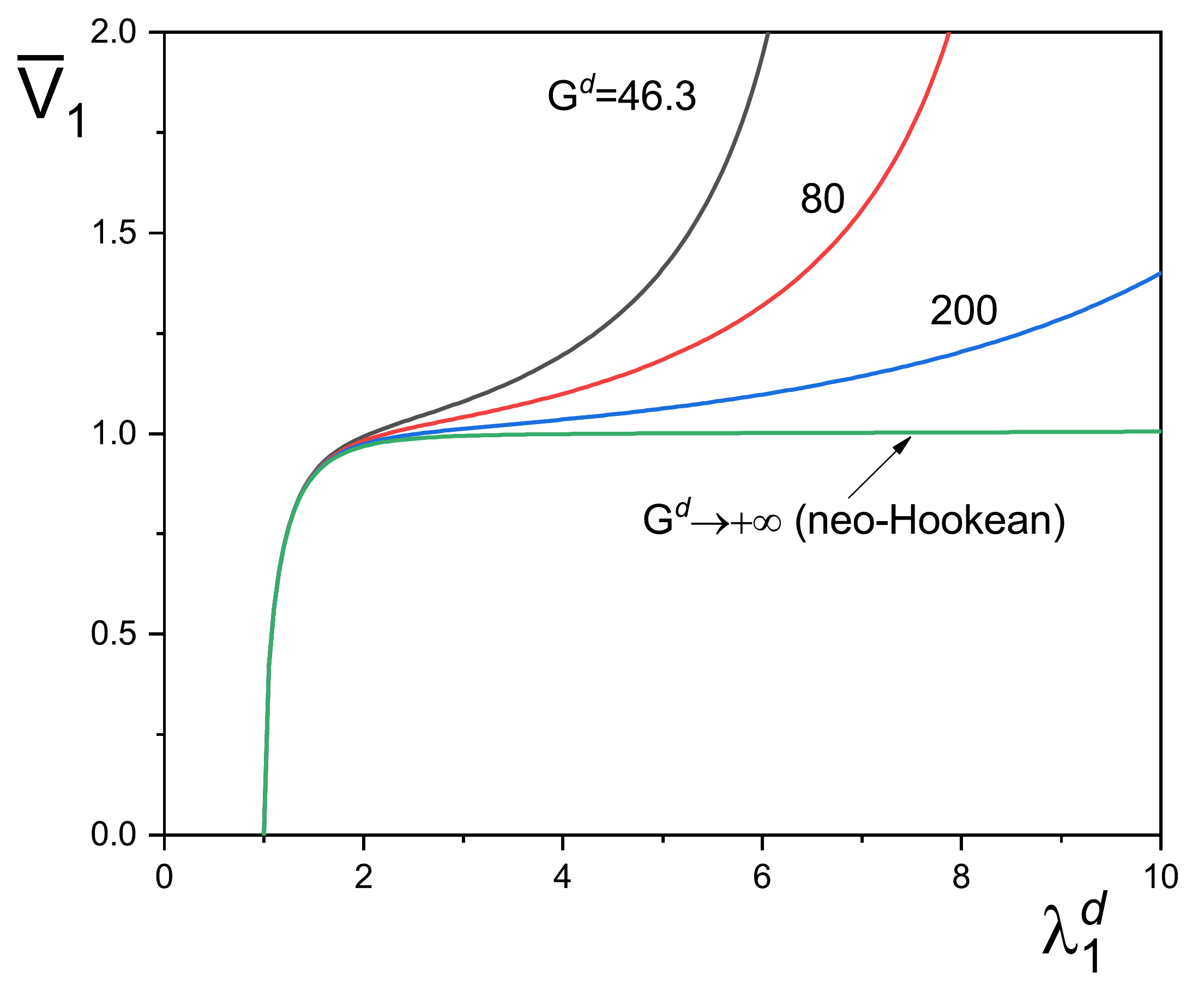}
\caption{
{\footnotesize
\color{black}
Nonlinear behaviors of Gent dielectric elastomers for varying $G^d$ subject to a voltage $\overline V_1$.
The Gent model degenerates to the neo-Hookean model in the limiting case $G^d\to \infty$.\color{black}
}
}
\label{neo-Hookean_Gent}
\end{figure}

\color{black}
Most polymers exhibit the strain stiffening effect, which can be well captured by the Gent model \citep{Gent1996}. In particular, the Gent model recovers the neo-Hookean model \citep{Rivlin1948} in a limiting case,  which describes solids with unlimited extensibility that do not exist in nature. See Figure \ref{neo-Hookean_Gent}  for a comparison between the Gent and neo-Hookean responses.
\color{black}

For our calculations we use the ideal Gent dielectric model \color{black}\citep{Zhao2007} \color{black} and the Gent elastic model to describe the dielectric and elastic elastomers, respectively, as
 \begin{align}\label{Gent1}
& W_1^{d}(\lambda^d_1,D_1)=-\frac{\mu^{d}G^{{d}}}{2}\ln \left[1-\frac{(\lambda_1^{d})^2+(\lambda_1^{d})^{-2}-2}{G^{d}}\right] + \frac{1}{2\varepsilon}(\lambda_1^{d})^{-2}D_1^2, \notag \\[6pt]
& W^{d}(\lambda^d,D_2)=-\frac{\mu^{d}G^{{d}}}{2}\ln\left[1-\frac{(\lambda^{d})^2+(\lambda^{d})^{-2}-2}{G^{d}}\right]+\frac{1}{2\varepsilon}(\lambda^{d})^{-2}D_2^2, \notag \\[6pt]
& W^{e}(\lambda^e)=-\frac{\mu^{e}G^{{e}}}{2}\ln\left[1-\frac{(\lambda^{e})^2+(\lambda^{e})^{-2}-2}{G^{e}}\right],
 \end{align} 
where $G^s$ is the dimensionless stiffening parameter of the $s$-th block. 
Throughout the paper, we use  data \color{black} as collected by \cite{Shmuel2017} for Silicone CF19-2186 \color{black} from the manufacturer Nusil$^\text{TM}$ (Nusil$^\text{TM}$ Technology LLC, Carpinterio, USA)  to model the dielectric elastomer, \color{black}with $\mu^d=333\text{kPa},\ \varepsilon^d=2.5\times10^{-11}\text{F/m},\ G^d=46.3$ and  dielectric strength $E^d_B=235\text{MV/m}$.
We set $G^e=46.3$ for the elastic block also. 
Then the normalized breakdown electric field of the dielectric elastomer is $\overline{E}^d_B=E^d_B\sqrt{\varepsilon^d/\mu^d}=2.04$. In the following calculations, the dimensionless actuation electric field of the dielectric elastomer is selected below this value to avoid electric breakdown failure.\color{black}

From Eqs. \eqref{voltage-displacement1} and \eqref{voltage1}, we determine the homogeneous stretch $\lambda^d_1$ of the  dielectric elastomer by solving the equation
\begin{equation}\label{stretch1}
\overline P=-\lambda^d_1 \overline V_1^2-\frac{G^d\left [(\lambda_1^d)^4-1\right ]}{\lambda_1^d-(G^d+2)(\lambda_1^d)^3+(\lambda_1^d)^5},
\end{equation}
once the dimensionless pre-stress $\overline P=P/\mu$ and voltage $\overline V_1=V_1\sqrt{\varepsilon/\mu^d}/H^d$ are prescribed. Since no mechanical loading is required to activate the bending deformation of the bilayer we have $\overline P=0$ in the following calculations. 

\color{black}
By solving Eq. \eqref{stretch1} with $P=0$, we obtain Figure \ref{neo-Hookean_Gent}.
It  displays the nonlinear  $\overline V_1-\lambda_1^d$ response of the dielectric elastomer during the homogeneous deformation. The graph shows that the stretch $\lambda_1^d$ increases monotonously with the increase of the applied voltage $\overline V_1$. Materials with finite $G^d$ clearly exhibit the strain stiffening effect, and the stretchability of the material increases as $G^d$ increases. Finally, note that the Gent model recovers the neo-Hookean model with infinite stretchability as $G^d\to\infty$. 
We note here that the \textit{pull-in instability} can be avoided for dielectric elastomers under plane strain deformation, as shown by \cite{Yang2017} and \cite{Su2019}.
\color{black} 

According to Eq. \eqref{normal-stress}, we obtain the stresses in the radial and hoop directions of the bilayer as
\begin{align}\label{normal-stress1}
&\overline \tau^d_{rr}=\frac{G^d}{2}\ln\left[\frac{G^d-\left(\lambda_a\right)^{-2}-\left(\lambda_a\right)^2+2}{G^d-\left(\lambda^d\right)^{-2}-\left(\lambda^d\right)^2+2}\right] +\frac{\overline D_2^2}{2}\left[\left(\lambda^d\right)^{-2}-\left(\lambda_a\right)^{-2}\right],\notag \\[6pt]
&\overline \tau^e_{rr}=\frac{G^e\overline\mu}{2}\ln\left[\frac{G^e-\left(\lambda_b\right)^{-2}-\left(\lambda_b\right)^2+2}{G^e-\left(\lambda^e\right)^{-2}-\left(\lambda^e\right)^2+2}\right],\notag \\[6pt]
&\overline \tau^d_{\theta\theta} = 
\frac{G^d\left[\left(\lambda^d\right)^4-1\right]}{\left(2+G^d\right)\left(\lambda^d\right)^2 - \left(\lambda^d\right)^4-1} 
+ \frac{G^d}{2}\ln \left[\frac{G^d-\left(\lambda_a\right)^{-2}-\left(\lambda_a\right)^2+2}{G^d-\left(\lambda^d\right)^{-2}-\left(\lambda^d\right)^2+2} \right] 
\notag \\
& \qquad\qquad - \frac{\overline D_2^2}{2}\left[\left(\lambda^d\right)^{-2}+\left(\lambda_a\right)^{-2}\right],
\notag\\[6pt]
&\overline \tau^e_{\theta\theta} 
= \frac{\overline\mu G^e\left[\left(\lambda^e\right)^4-1\right]}{\left(2+G^e\right)\left(\lambda^e\right)^2-\left(\lambda^e\right)^4-1} + \frac{\overline \mu G^e}{2}\ln \left[\frac{G^e-\left(\lambda_b\right)^{-2}-\left(\lambda_b\right)^2+2}{G^e-\left(\lambda^e\right)^{-2}-\left(\lambda^e\right)^2+2}\right].
\end{align}
Then we determine $D_2$ with the connection \eqref{connection} as
\begin{equation}\label{electric-displacement}
\overline D_2=\frac{\overline V_2 (\lambda^d_1)^2}{2}\left[\left(\lambda^e_m\right)^2-\left(\frac{\lambda_a}{\lambda^d_1}\right)^2\right]\left(\ln\frac{\lambda^d_m}{\lambda_a}\right)^{-1},
\end{equation}
and the interfacial condition \eqref{interfacial-stretch} reads
\begin{multline}\label{interfacial1}
G^d \ln \left[\frac{G^d-\left(\lambda_a\right)^{-2}-\left(\lambda_a\right)^2+2}{G^d-\left(\lambda^d_m\right)^{-2}-\left(\lambda^d_m\right)^2+2} \right] + \overline D_2^2\left[\left(\lambda^d_m\right)^{-2}-\left(\lambda_a\right)^{-2}\right]\\
 = \overline\mu G^e\ln\left[\frac{G^e-\left(\lambda_b\right)^{-2}-\left(\lambda_b\right)^2+2}{G^e-\left(\lambda^e_m\right)^{-2}-\left(\lambda^e_m\right)^2+2}\right],
\end{multline}
where $\overline \tau_{ii}=\tau_{ii}/\mu_d \ (i=r,\theta)$, $\overline V_2=V_2\sqrt{\varepsilon/\mu^d}/H^d$ and $\overline D_2=D_2/\sqrt{\mu^d\varepsilon}$ are the dimensionless measures of $\tau_{ii}$, $V_2$ and $D_2$, respectively, and $\overline \mu=\mu^e/\mu^d$ is the ratio of the shear moduli of the elastic and dielectric elastomers.

By using $\lambda^d=\lambda^d_1\omega r$ and $\lambda^e=\omega r$, Eq. \eqref {interfacial-moment} can be rewritten as
\begin{equation}\label{zeromoment}
\left(\lambda^d_1\right)^{-2}\int_{\lambda_a}^{\lambda_m^d}\overline\tau_{\theta\theta}^d \lambda^d\,\text d\lambda^d+\overline \mu\int_{\lambda_m^e}^{\lambda_b}\overline\tau_{\theta\theta}^e \lambda^e\,\text d\lambda^e=0.
\end{equation}
The bending deformation of the bilayer can be completely determined by solving Eqs. \eqref{interfacial2} and \eqref{stretch1}-\eqref{zeromoment}, once the loadings $\overline P$, $\overline V_1$ and $\overline V_2$ are prescribed. 

For the considered Gent elastomer with corresponding deformation gradient \eqref{total-deformation}, we obtain the following components of the  electro-elastic moduli tensors (with $s$ omitted)
\begin{align}
& \mathcal A_{1111}=\mu\left(2\lambda^{-2}\overline W_1+4\lambda^{-4}\overline W_{11}+\lambda^{-2}\overline D_2\right), 
&&\mathcal A_{1122}=4\mu\overline W_{11}, 
\notag \\
& \mathcal A_{2222}=2\mu\lambda^{2}(\overline W_1+2\lambda^2\overline W_{11}), 
&& \mathcal A_{1221}=0, 
\notag \\
&\mathcal A_{1212}=\mu\lambda^{-2}(2\overline W_1+\overline D_2^2),
&& \mathcal A_{2121}=2\mu\lambda^{2}\overline W_1,
\notag \\ 
&\Gamma_{122}=\lambda^{-1}\overline D_2\sqrt{\mu\varepsilon^{-1}}, 
&& K_{22}=\varepsilon^{-1},
\end{align}
where
\begin{equation}
\overline W_1= \frac{G\lambda^2}{2\left[(G+2)\lambda^2-\lambda^{4}-1)\right]},
\qquad
\overline W_{11}=\frac{G\lambda^4}{2\left[1 - (G+2)\lambda^2+\lambda^4\right]^2}.
\end{equation}


\subsection{Bending response of the bilayer}


\begin{figure}[h!]
\centering
\includegraphics[width=0.75\textwidth]{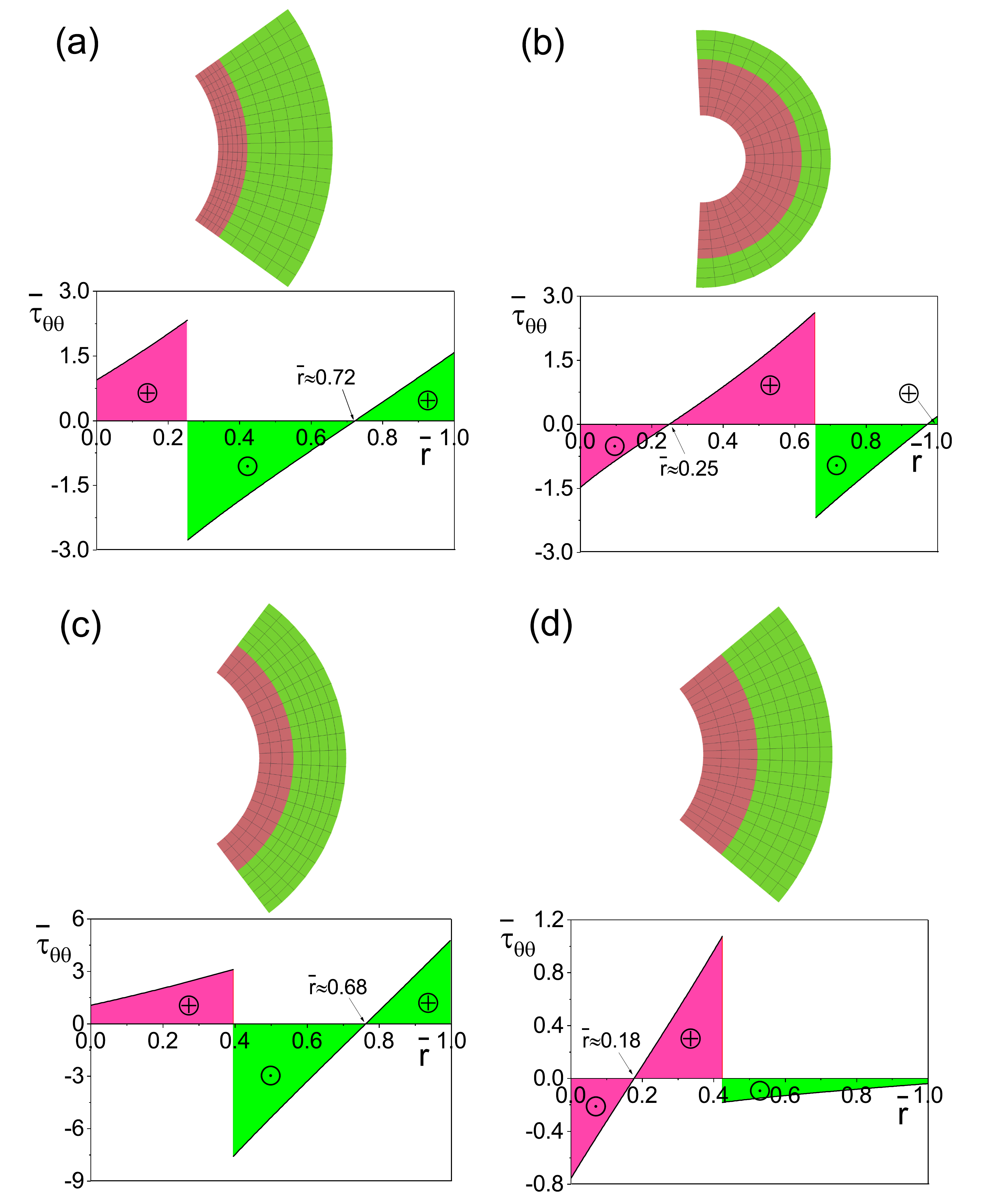}
\caption{
{\footnotesize
Bending deformations of dielectric-elastomeric bilayers for fixed $L^d/H^d=3$ and $G^d=G^e=46.3$ and various thickness and shear modulus ratios in response to $\overline V_1=1, \overline V_2=\overline P=0$: (a) $\overline H=2, \overline \mu=$2.5; (b) $\overline H=0.4, \overline \mu=$2.5; (c) $\overline H=1, \overline \mu=$10; (d) $\overline H=1, \overline \mu=$0.1. The red and green phases represent the dielectric and elastic blocks, respectively. The top and bottom rows of each case correspond to the self-equilibrium shape and the associated hoop stress distribution of the bilayer, respectively.
}
}
\label{figurex}
\end{figure}

In Figure \ref{figurex} we display the self-equilibrium shapes and the associated hoop stress distributions along the thickness of dielectric-elastomeric bilayers for various thickness and shear modulus ratios. 

We define $\overline r=(r-r_a)/(r_b-r_a)$ as the normalized radial position of a point in the bilayer. 
We first apply a constant voltage $\overline V_1=1$ to stretch the dielectric elastomer, and then take $\overline V_2=0$ to completely release it once the dielectric and elastic elastomers are perfectly bonded. 

We can see that the physical parameters have significant influences on the bending response of the bilayer. The circumferential stresses of the two elastomers are discontinuous at the interface, which may result in slide, exfoliation or crack during the bending deformation of the bilayer thus should be avoided in film-substrate systems \citep{Suo1995, Zhao2014, Vella2009}. Both the dielectric and elastic elastomers can be finely designed to be transversely compressed or stretched, and the neutral axis which corresponds to vanishing circumferential stress can appear in the dielectric layer (case $d$), the elastic layer (cases $a, c$) or both the dielectric and elastic layers (case $b$). 
These results may provide guidance to design stretchable/flexible electronic devices where the active devices are placed on the neutral axis to avoid failure \citep{Morimoto2015}.

\begin{figure}[h!]
\centering
\includegraphics[width=0.6\textwidth]{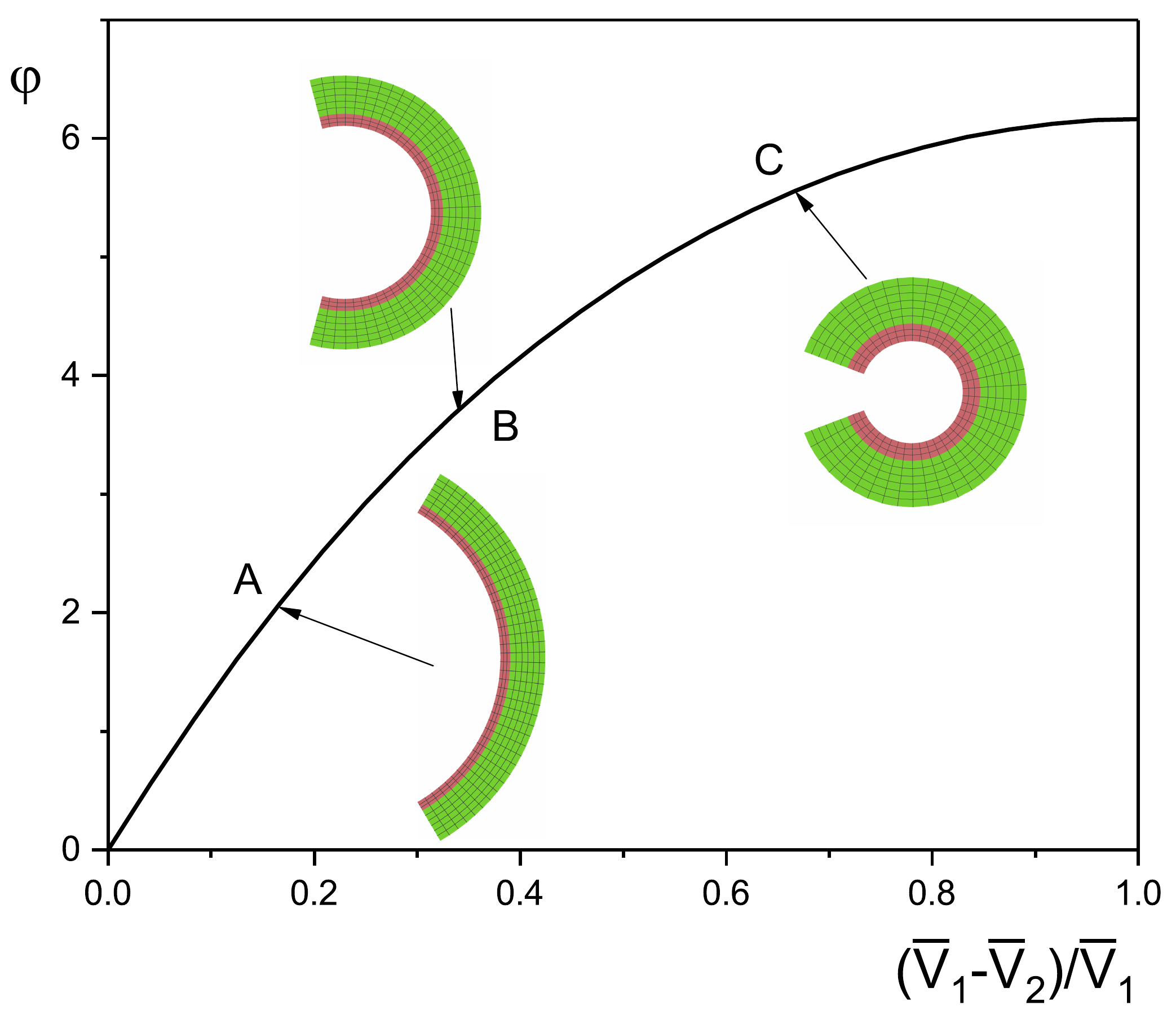}
\caption{
{\footnotesize
Plot of bending angle $\varphi$ as the function of voltage change $(\overline V_1-\overline V_2)/\overline V_1$ for a dielectric-elastic bilayer with $L^d/H^d=3, H^e/H^d=1, \mu^e/\mu^d=10$ and $G^d=G^e=46.3$. The applied voltage to stretch the dielectric block is $\overline V_1=1.2$ and then is reduced gradually to zero. 
For the bilayers A, B and C, we have $\overline V_2=$1, 0.8 and 0.4 respectively.
}
}
\label{figure3}
\end{figure}
The effect of the applied voltage on the bending of the dielectric-elastic bilayer is presented in Figure \ref{figure3}. 
We can see that the bending angle $\varphi$ increases gradually as $(\overline V_1-\overline V_2)/\overline V_1$ increases, i.e., the bilayer curls up as the applied voltage $\overline V_1$ reduces. 
In the calculation no mechanical loading is applied ($P=0$), and we conclude that by solely tuning the applied voltage, the smart bending deformation of the bilayer can be designed to achieve  desired motions, such as those presented in Figure \ref{figure1}.

\begin{figure}[h!]
\centering
\includegraphics[width=0.6\textwidth]{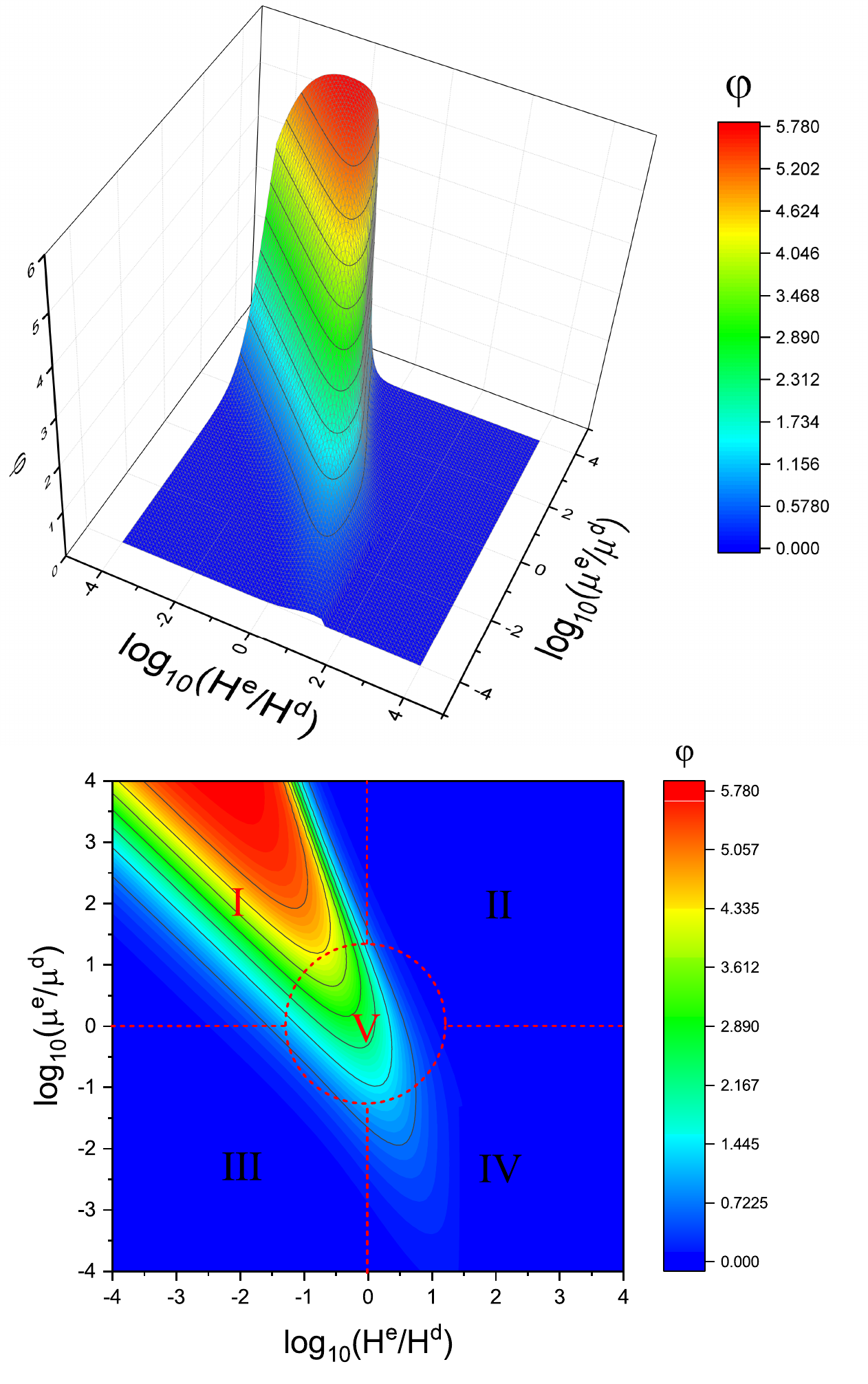}
\caption{
{\footnotesize
Plot of bending angle $\varphi$ as the functions of $\log_{10}(H^e/H^d)$ and $\log_{10}(\mu^e/\mu^d)$ for a dielectric-elastic bilayer in the case where $L^d/H^d=3$ and $G^d=G^m=46.3$ (top) and its top view version (bottom). The applied voltage to initially stretch the dielectric block is $\overline V_1=0.8$ and then is completely released ($\overline V_2=0$).
}
}
\label{figure4}
\end{figure}

Figure \ref{figure4} displays the effects of the relative parameters of the two elastomers $H^e/H^d$ and $\mu^d/\mu^e$ on the bending angle $\varphi$ of a bilayer with $L^d/H^d=3$ and $G^d=G^e=46.3$. 
We first apply a voltage $\overline V_1=0.8$ to stretch the dielectric block and then release the voltage once the two blocks are perfectly bonded ($\overline V_2=0$). 
It can be seen that the bending behavior of the bilayer is significantly influenced by the related parameters. 

The physical mechanism behind the bending response of the bilayer can be understood as follows.
If the dielectric block was by itself, it would shrink homogeneously in length to its initial configuration once the voltage $\overline V_1$ is released. 
The bonded elastic block prevents the recovery of the original dielectric block geometry, and results in an inhomogeneous deformation which triggers the bending deformation of the bilayer. 
If the constraint exerted by the elastic block on the deformation of the dielectric block is weak (e.g., the elastic block is thin and soft compared with the dielectric block as shown in Area \uppercase\expandafter{\romannumeral3} in Figure\ref{figure4}), then the dielectric block deforms almost in its expected homogeneous way, and the bending of the bilayer can be almost ignored. 
On the other hand, if the elastic block is  stiff and thick (Area \uppercase\expandafter{\romannumeral2}) or soft but extremely thick (Area \uppercase\expandafter{\romannumeral4}), then the constraint exerted by the elastic block on the deformation of the dielectric block is so strong that the dielectric block almost keeps the same length as the elastic block, once  they are bonded together.
To achieve a large bending deformation, the shear modulus and the thickness of the dielectric phase of the bilayer should be of the same order of magnitude as those of the elastic phase (Area \uppercase\expandafter{\romannumeral5}), otherwise the dielectric phase of the bilayer should be soft but thick compared with the elastic phase (Area \uppercase\expandafter{\romannumeral1}).

\subsection{Instability analysis}
\begin{figure}[h!]
\centering
\includegraphics[width=0.9\textwidth]{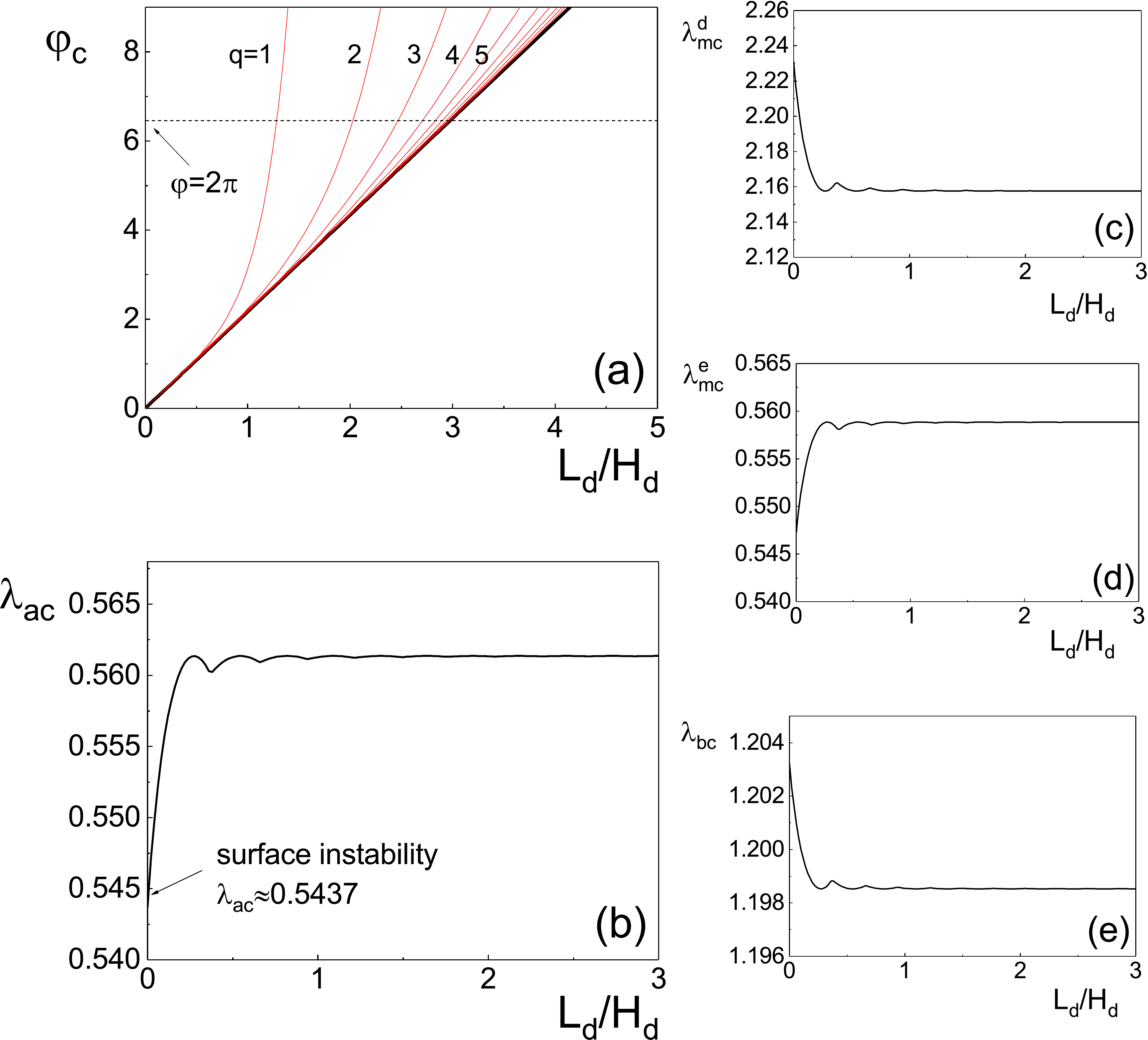}
\caption{
{\footnotesize
Bending instability of a neo-Hookean ($G^d=G^e=10^4$) dielectric-elastomeric bilayer with $\overline H=\overline \mu=1$: the critical bending angle $\varphi_c$ and the critical stretches $\lambda_{ac}, \lambda^d_{mc}, \lambda^e_{mc}$ and $\lambda_{bc}$ at the onset of buckling versus the width aspect ratio of the dielectric elastomer $L^d/H^d$ are shown in (a)-(e), respectively.
}
}
\label{figurey}
\end{figure}

\begin{figure}[h!]
\centering
\includegraphics[width=0.8\textwidth]{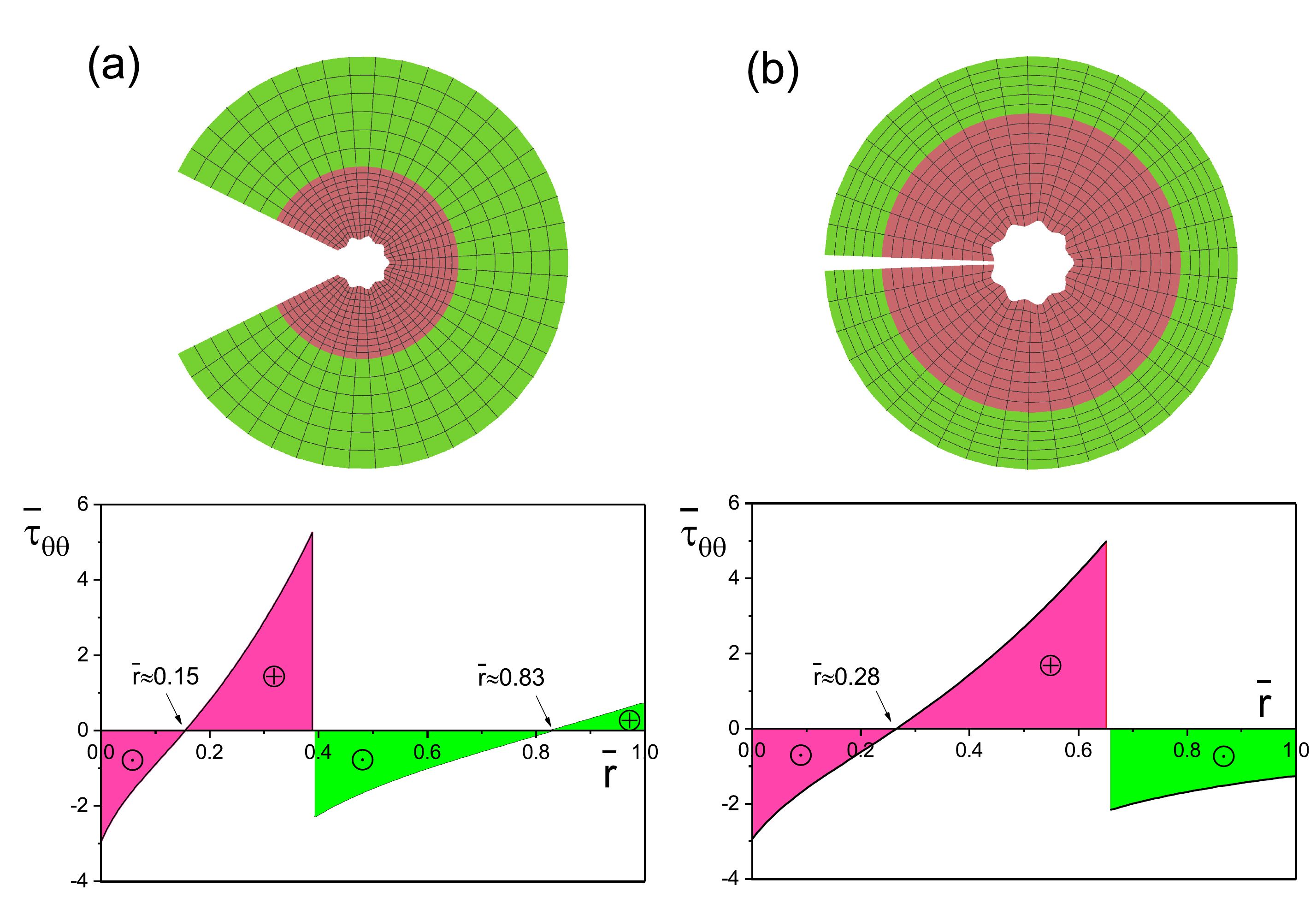}
\caption{
{\footnotesize
Bending instability of  Gent dielectric-elastic bilayers with $ G^d=G^e=46.3$ and (a): $H^e/H^d=1$, $\mu^e/\mu^d=1$, $L^d/H^d=3$ and (b): $H^e/H^d=0.2$, $\mu^e/\mu^d=0.5$, $L^d/H^d=3$. The top row shows the wrinkling shapes when instability occurs, with 8 (Case (a)) and 9 (Case (b)) wrinkles occurring on the outer surface of the bilayer, as the applied voltage $\overline V_1$ reaches 1.17 and 1.38, respectively. The bottom row shows the distributions of the corresponding circumferential stresses in the bilayers.
}
}
\label{figure5}
\end{figure}

For the stability analysis, we focus on the scenario where the voltage  is completely removed from the bilayer ($\overline V_2=0$), and find out how much it can be bent until wrinkles appear. 

In Figure \ref{figurey} we compare the results of our model with those of the purely elastic problem. 
We take $G^d=G^e=10^4$ to specialise the analysis to neo-Hookean materials \citep{Destrade09, Su2018}, and set $\overline H=\overline \mu=1$. 

Figure \ref{figurey}(a) displays the critical bending angle $\varphi_{c}$ as 
a function of the aspect ratio of the dielectric elastomer $L^d/H^d$. 
By taking the number of wrinkles in turn as $q=1,2,3,...$ and solving the dispersive equation \eqref{dispersion}, we  obtain a series of $\overline V_{1c}-L^d/H^d$ curves (in fact, there are each a scaled version of the $q=1$ curve, see \cite{Destrade09}). 
Note that only the highest points located on the highlighted black bold curve are meaningful, as they indicate the earliest onset of instability. 
From this plot, the critical bending angle $\varphi_c$ and the number of wrinkles $q$ of the buckled bilayer can be predicted once $L^d/H^d$ is given. 
We can see that the critical bending angle $\varphi_c$ increases linearly as $L^d/H^d$ increases, and that bilayers with  $L^d/H^d>3$ can be bent into closed tubes without buckling. 

The variations of the associated critical stretches $\lambda_{ac}$, $\lambda^d_{mc}$, $\lambda^e_{mc}$ and $\lambda_{bc}$ with $L^d/H^d$ are displayed in Figures \ref{figurey}(a)-(e), respectively, with only the meaningful curves being presented. 
When $L^d/H^d \to 0$, \color{black}the threshold of the bending instability approaches to the critical stretch $\lambda_{ac}=0.5437$ at which an incompressible neo-Hookean half-space under plane strain becomes unstable \citep{Biot1963}.\color{black} 

Results presented in Figure \ref{figure4} show that to obtain a considerable bending deformation, the properties of the bilayer should lie in Area \uppercase\expandafter{\romannumeral1} or  Area \uppercase\expandafter{\romannumeral5}.
We first consider the bilayers from Area \uppercase\expandafter{\romannumeral5}. 
We can see from Figure \ref{figure5} that for these bilayers,  the inner part of the dielectric block is always compressed when buckling happens, and wrinkles appear on the inner surface of the dielectric block. 
The outer part of the elastic block can be either in tension (Figure \ref{figure5}(a)) or in compression (Figure \ref{figure5}(b)).
The elastic elastomer is sufficiently soft and thick to sustain compression and wrinkles do not  appear on its compressed face (when there is one). 
Notice that the perturbation decays rapidly along the radius, and that the displacement on the inner face is several orders of magnitude larger than that on the outer face. The material properties of the the bilayer can be properly designed so that it can be bent into a closed tube ($\varphi_c>2 \pi$) without encountering buckling (not shown here). 

\begin{figure}[h!]
\centering
\includegraphics[width=0.8\textwidth]{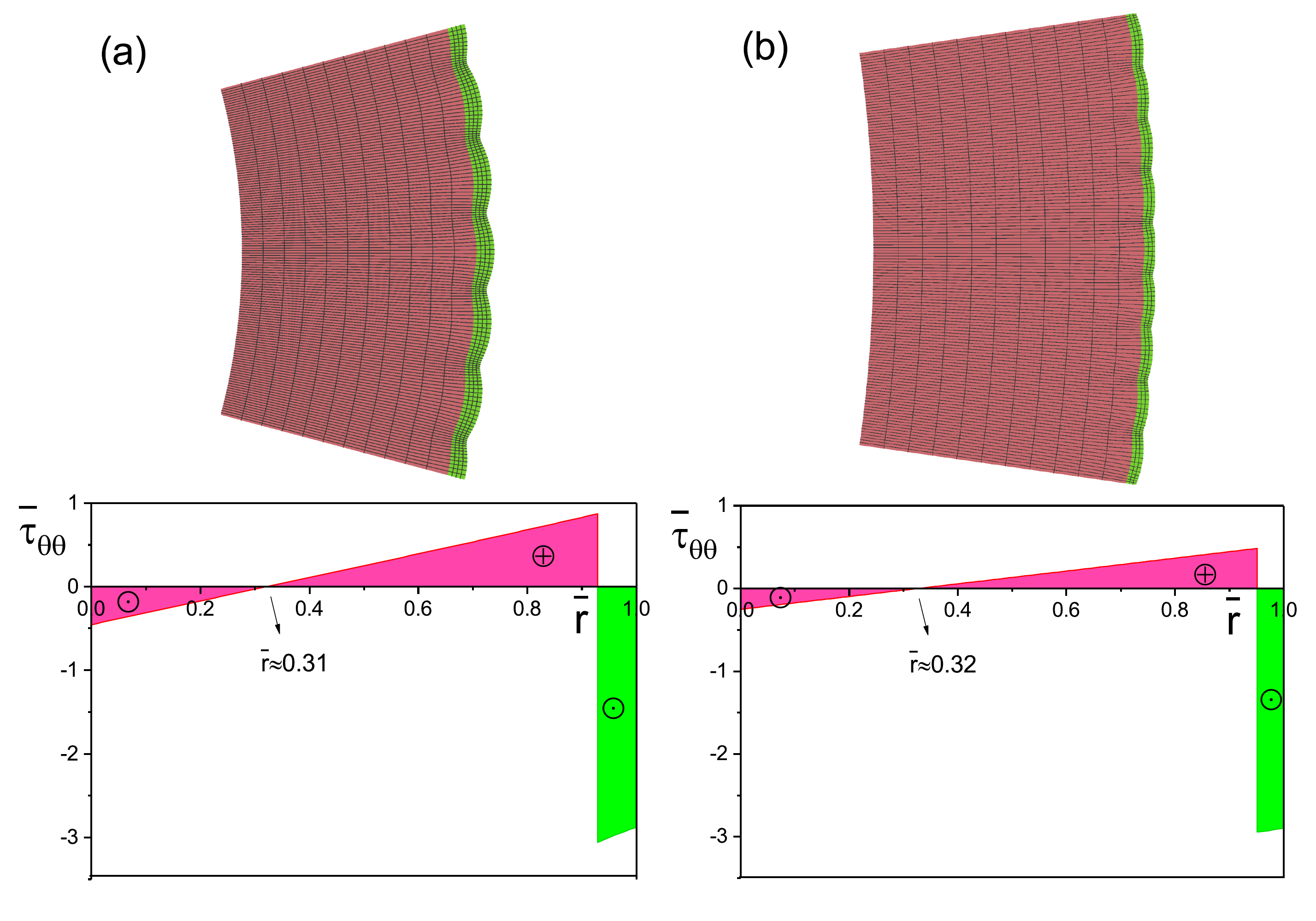}
\caption{
{\footnotesize
Elastic bending instability of a dielectric-elastic bilayer for fixed $ G^d=G^e=46.3$ and various phase properties: (a) $H^e/H^d=0.05, \mu^e/\mu^d=2.5, L^d/H^d=2$; (b) $H^e/H^d=0.03, \mu^e/\mu^d=2.5, L^d/H^d=2$. The top row shows the wrinkling shapes when instability occurs, with 6 and 9 wrinkles occur on the outer surface of the bilayer, as the applied voltages $\overline V_1$ reach 0.94 and 0.9, respectively. The bottom row shows the distributions of the corresponding circumferential stresses in the bilayers.
}
}
\label{figure6}
\end{figure}

Next we consider instabilities of the bilayers from Area \uppercase\expandafter{\romannumeral1}, with results presented in Figure \ref{figure6}. 
Here the elastic layer of the solid is always compressed, and because it is relatively stiff and thin compared with the dielectric layer, it cannot sustain compression and wrinkles appear. 
In this case, the bending deformation is relatively inconspicuous, and for the case $H^e/H^d\to 0$, the problem of instability of a layer resting on a substrate is recovered \citep{Huang2005, Yin2018}. 

We observe that although theoretically the bilayers from Area \uppercase\expandafter{\romannumeral1} possess more favorable ``bendability'' than those from Area \uppercase\expandafter{\romannumeral5}, they are more susceptible to fail by the wrinkling instability.

\section{Conclusion and discussion}
\label{section5}


In conclusion, we have investigated the bending response of a bilayer composed of a dielectric block and an elastic block in response to a voltage, and the associated buckling phenomenon, based on nonlinear theory of electro{\color{black}{-}}elasticity and  linearized incremental theory. 

We first derived the governing equations of the static bending response of the bilayer for a general form of energy function, and obtained analytical solutions for Gent materials. 
We found that the bending angle of the bilayer can be tuned by the application of voltage and the relative physical parameters of the dielectric and elastic elastomers, and we revealed the mechanism required to obtain a considerable smart bending of the bilayer. 

We  recast the associated incremental problem in the form of first-order differential systems. 
We then used the surface impedance matrix method to build a robust numerical procedure for obtaining and solving the dispersion equation which predicts the critical condition when the bending instability occurs in the bilayer. 
We conducted calculations for Gent materials and recovered the results of the purely elastic problem. 
We showed that the pattern formation of a buckled bilayer can be selected by finely tuning the physical properties of the bilayer and the applied voltage.

\color{black}
The assumption of \emph{plane strain deformation}  was adopted to simplify the theoretical model of the problem by taking the in-plane width of each elastomer much longer than its thickness and length, but  our analysis can easily be extended to the case of an arbitrary tri-axial pre-stretch \citep{Haughton99}.

\emph{Viscoelasticity} has been experimentally observed to have an effect on failure modes of dielectric elastomers \citep{Plante2006}. To study its influence, a nonlinear field theory of viscoelastic electroelasticity \citep{Hong2011} should be employed, which is beyond the scope of the current paper.
\color{black}

The results obtained here may provide references for  the design and fabrication of dielectric-elastic bilayers that are capable of \emph{finite smart bending}, and can be generalized to the case of smart bending activated by pH, temperature or light signals.
\color{black}They can also provide a basis for a \emph{post-buckling analysis}, be it theoretical or numerical (Finite Element Method). 
As shown experimentally by \cite{Gent99}, creases precede wrinkles on the inner face of bent rubber blocks. 
Indeed,  a post-buckling analysis of compressed homogeneous half spaces reveals that creases are subcritical.
However, when the half-space is coated with a layer of arbitrary thickness, creases can be supercritical when the layer is at least 1.74 times stiffer than the substrate \citep{Cai99}, in which case wrinkles do occur at the predicted linearised criterion and remain stable for a consequent compression \citep{Cao12}.

\color{black}



\section*{Acknowledgments}


This work was supported by a Government of Ireland Postdoctoral Fellowship from the Irish Research Council (No. GOIPD/2017/1208) and by the National Natural Science Foundation of China (No. 11621062).
WQC and YPS acknowledge the support from the Shenzhen Scientific and Technological Fund for R$\&$D (No. JCYJ20170816172316775). 
\color{black}
BW acknowledges the awarding of Research Fellow at Politecnico di Torino.
\color{black}
MD thanks Zhejiang University for funding research visits  to Hangzhou. YPS is also grateful to Chengjun Wang and Renwei Mao at Zhejiang University for fruitful discussions.







\begin{thebibliography}{99}


\bibitem[Bertoldi and Gei(2011)]{Bertoldi2011}
Bertoldi, K., Gei, M., 2011. 
Instabilities in multilayered soft dielectrics. 
J. Mech. Phys. Solids 59, 18-42.

\bibitem[Bigoni(2012)]{Bigoni12}
Bigoni, D., 2012. 
\emph{Nonlinear solid mechanics: Bifurcation theory and material instability}.  
University Press, Cambridge.

\bibitem[Biot(1963)]{Biot1963}
Biot, M.A., 1963. 
Surface instability of rubber in compression. 
Appl. Sci. Res. A 12, 168-182.

\bibitem[Bortot and Shmuel(2018)]{Shmuel2018}
Bortot, E., Shmuel, G., 2018. 
Prismatic bifurcations of soft dielectric tubes. 
Int. J. Eng. Sci. 124, 104-114.

\bibitem[Brochu and Pei(2010)]{Brochu2010}
Brochu, P., Pei, Q., 2010. 
Advances in dielectric elastomers for actuators and artificial muscles. 
Macromol. Rapid Comm. 31(1), 10-36.

\color{black}
\bibitem[Cai and Fu(1999)]{Cai99}
Cai, Z.X., Fu, Y.B., 1999.
On the imperfection sensitivity of a coated elastic half-space.
Proc. R. Soc. Lond. A  455, 3285-3309.

\bibitem[Cao and Hutchinson(2012)]{Cao12}
Cao, Y.P., Hutchinson, J.W., 2012.
From wrinkles to creases in elastomers: the in-stability and imperfection-sensitivity of wrinkling.
 Proc. R. Soc. A 468, 94–115
\color{black}

\bibitem[Destrade et al.(2009)]{Destrade09}
Destrade, M., N\'i Annaidh, A., Coman, C.D., 2009. 
Bending instabilities of soft biological tissues. 
Int. J. Solids Struct. 46(25-26), 4322-4330.

\bibitem[Dorfmann and Ogden(2005)]{Dorf05}
Dorfmann, A., Ogden, R.W., 2005. 
Nonlinear electroelasticity. 
Acta Mech. 174(3-4), 167-183.

\bibitem[Dorfmann and Ogden(2006)]{Dorf06}
Dorfmann, A., Ogden, R.W., 2006. 
Nonlinear electroelastic deformations. 
J. Elasticity 82(2), 99-127.

\bibitem[Dorfmann and Ogden(2010)]{Dorf10}
Dorfmann, A., Ogden, R.W., 2010. 
Nonlinear electroelastostatics: Incremental equations and stability. 
Int. J. Eng. Sci. 48(1), 1-14.

\bibitem[Du et al.(2018)]{Du2018}
Du, Y.K., L$\ddot {\text u}$, C.F., Chen, W.Q., Destrade, M., 2018. 
Modified multiplicative decomposition model for tissue growth: Beyond the initial stress-free state. 
J. Mech. Phys. Solids 118, 133-151.

\bibitem[Fu et al.(2018)]{Fu2018}
Fu, H.R., Nan, K.W., Bai, W.B., Huang, W., Bai, K., Lu, L.Y., Zhou, C.Q., Liu, Y.P., Liu, F., Wang, J.T., Han, M.D., Yan, Z., Luan, H.W., Zhang, Y.T., Zhao, J.N., Cheng, X., Li, M.Y., Lee, J.W., Liu, Y., Fang, D.N., Li, X.L., Huang, Y.G., Zhang, Y.H., Rogers, J., 2018. 
Morphable 3D mesostructures and microelectronic devices by multistable buckling mechanics. 
Nat. Mater. 17(3), 268.

\bibitem[Gent(1996)]{Gent1996}
Gent, A.N., 1996. 
A new constitutive relation for rubber. 
Rubber Chem. Technol. 69(1), 59-61.

\bibitem[Gent and Cho(1999)]{Gent99}
Gent, A.N., Cho, I.S., 1999. 
Surface instabilities in compressed or bent rubber blocks. 
Rubber Chem. Technol. 72(2), 253-262.

\color{black}
\bibitem[Getz et al.(2017)]{Shmuel2017}
Getz, R., Kochmann, D.M., Shmuel, G., 2017. 
Voltage-controlled complete stopbands in two-dimensional soft dielectrics. 
Int. J. Solids Struct. 113, 24-36.
\color{black}

\bibitem[Goshkoderia and Rudykh(2017)]{Rudykh2017}
Goshkoderia, A., Rudykh, S., 2017. 
Electromechanical macroscopic instabilities in soft dielectric elastomer composites with periodic microstructures.
Eur. J. Mech.-A/Solid. 65, 243-256.

\bibitem[Green and Zerna(1954)]{Green54}
Green, A.E., Zerna, W., 1954. 
\emph{Theoretical Elasticity}. 
University Press, Oxford. Reprinted by Dover, New York.

\color{black}
\bibitem[Haughton(1999)]{Haughton99}
Haughton, D.M., 1999. 
Flexure and compression of incompressible elastic plates.
Int. J. Eng. Sci. 37, 1693–1708.
\color{black}

\bibitem[He et al.(2017)]{Wang2017}
He, L.W., Lou, J., Du, J.K., Wang, J., 2017. 
Finite bending of a dielectric elastomer actuator and pre-stretch effects. 
Int. J. Mech. Sci. 122, 120-128.

\color{black}
\bibitem[Hong(2011)]{Hong2011}
Hong, W., 2011. 
Modeling viscoelastic dielectrics. J. Mech. Phys. Solids 59(3), 637-650.
\color{black}

\bibitem[Huang et al.(2005)]{Huang2005}
Huang, Z.Y., Hong, W., Suo, Z.G., 2005. 
Nonlinear analyses of wrinkles in a film bonded to a compliant substrate. 
J. Mech. Phys. Solids 53(9), 2101-2118.

\bibitem[Li et al.(2015)]{Li2015}
Li, Q.W., Liu, C.H., Lin, Y.H., Liu, L., Jiang, K.L., Fan, S.S., 2015. 
Large-strain, multiform movements from designable electrothermal actuators based on large highly anisotropic carbon nanotube sheets. 
ACS Nano 9(1), 409-418.


\bibitem[Li et al.(2017)]{Li2017}
Li, T.F., Li, G.R., Liang, Y.M., Cheng, T.Y., Dai, J., Yang, X.X., Liu, B.Y., Zeng, Z.D., Huang, Z.L., Luo, Y.W., Xie, T., Yang, W., 2017. 
Fast-moving soft electronic fish. 
Sci. Adv. 3(4), e1602045.

\bibitem[Li et al.(2018)]{Li2018}
Li, T.F., Zou, Z.N., Mao, G.Y., Yang, X.X., Liang, Y.M., Li, C., Qu, S.X., Yang, W., 2018. 
Agile and resilient insect-scale robot. 
Soft Robot.

\bibitem[Liu et al.(2016)]{Liu2016}
Liu, Y., Yan, Z., Lin, Q., Guo, X.L., Han, M.D., Nan, K.W., Hwang, K.C., Huang, Y.G., Zhang, Y.H., Rogers, J.A., 2016. 
Guided formation of 3D helical mesostructures by mechanical buckling: Analytical modeling and experimental validation. 
Adv. Funct. Mater. 26(17), 2909-2918.

\bibitem[Madsen et al.(2006)]{Madsen2006}
Madsen, H.O., Krenk, S., Lind, N.C., 2006. 
Methods of structural safety. Courier Corporation.

\bibitem[Morimoto and Ashida(2015)]{Morimoto2015}
Morimoto, T.,  Ashida, F., 2015. 
Temperature-responsive bending of a bilayer gel. 
Int. J. Solids Struct. 56, 20-28.

\bibitem[Nardinocchi and Puntel(2017)]{Nardinocchi2017}
Nardinocchi, P., Puntel, E., 2017.
Swelling-induced wrinkling in layered gel beams. 
Proc. R. Soc. A 473(2207), 20170454.

\bibitem[O'Halloran et al.(2008)]{O'Halloran2008}
O'Halloran, A., O'Malley, F., McHugh, P., 2008. 
A review on dielectric elastomer actuators, technology, applications, and challenges. 
J. Appl. Phys. 104(7), 9.

\bibitem[Ogden(1997)]{Ogden97}
Ogden, R.W., 1997. 
\emph{Non-linear elastic deformations}. Dover, New York.

\bibitem[Pelrine et al.(2000)]{Pelrine2000}
Pelrine, R., Kornbluh, R., Pei, Q., Joseph, J., 2000. 
High-speed electrically actuated elastomers with strain greater than 100\%. 
Science 287(5454), 836-839.

\color{black}
\bibitem[Plante et al.(2006)]{Plante2006}
Plante, J.S., Dubowsky, S., 2006. 
Large-scale failure modes of dielectric elastomer actuators. Int. J. Solids Struct. 43(25-26), 7727-7751.
\color{black}

\bibitem[Roccabianca et al.(2010)]{Rocca10}
Roccabianca, S., Gei, M., Bigoni, D., 2010. 
Plane strain bifurcations of elastic layered structures subject to finite bending: 
Theory versus experiments. 
IMA J. Appl. Math. 75(4), 525-548.

\bibitem[Roccabianca et al.(2011)]{Rocca11}
Roccabianca, S., Bigoni, D., Gei, M., 2011. 
Long wavelength bifurcations and multiple neutral axes of elastic layered structures subject to finite bending. 
J. Mech. Mater. Struct. 6(1), 511-527.
\color{black}
\bibitem[Rivlin(1948)]{Rivlin1948}
Rivlin, R.S., 1948. 
Large elastic deformations of isotropic materials. I. Fundamental concepts. 
Philos. T. R. Soc. Lond. 240(822), 459-490.

\bibitem[Rivlin(1949)]{Rivlin1949}
Rivlin, R.S., 1949. 
Large elastic deformations of isotropic materials V: the problem of flexure. 
Proc. Roy. Soc. Lond. Ser. A 195, 463-473.
\color{black}

\bibitem[Su et al.(2018)]{Su2018b}
Su, Y.P., Broderick, H.C., Chen, W.Q., Destrade, M., 2018. 
Wrinkles in soft dielectric plates. 
J. Mech. Phys. Solids 119, 298-318.

\color{black}\bibitem[Su et al.(2019a)]{Su2019}
Su, Y.P., Chen, W.Q., Destrade, M., 2019a. 
Tuning the pull-in instability of soft dielectric elastomers through loading protocols. 
Int. J. Non-Lin. Mech. 113, 62-66.\color{black}

\bibitem[Su et al.(2019b)]{Su2018}
Su, Y.P., Wu, B., Chen, W.Q., Destrade, M., 2019b. 
Finite bending and pattern evolution of the associated instability for a dielectric elastomer slab. 
Int. J. Solids Struct. 158, 191-209.

\bibitem[Su et al.(2016)]{Su2016}
Su, Y.P., Zhou, W.J., Chen, W.Q., L$\ddot {\text u}$, C.F., 2016. 
On buckling of a soft incompressible electroactive hollow cylinder. 
Int. J. Solids Struct. (97-98), 400-416.

\bibitem[Suo(1995)]{Suo1995}
Suo, Z.G., 1995. 
Wrinkling of the oxide scale on an aluminum-containing alloy at high temperatures. 
J. Mech. Phys. Solids 43(6), 829-846.

\bibitem[Suo et al.(2008)]{Suo08}
Suo, Z.G., Zhao, X.H., Greene, W.H., 2008. 
A nonlinear field theory of deformable dielectrics. 
J. Mech. Phys. Solids 56(2), 467-486.

\bibitem[Vella et al.(2009)]{Vella2009}
Vella, D., Bico, J., Boudaoud, A., Roman, B., Reis, P. M., 2009. 
The macroscopic delamination of thin films from elastic substrates. 
P. Natl. A. Sci. 106(27), 10901-10906.

\bibitem[Wang et al.(2018)]{Wang2018}
Wang, C.J., Sim, K., Chen, J., Kim, H.J., Rao, Z., Li, Y.H., Chen, W.Q., Song, J.Z., Verduzco, R., Yu, C.J., 2018. 
Soft ultrathin electronics innervated adaptive fully soft robots. 
Adv. Mater. 30(13), 1706695.

\bibitem[Wang and Zhao(2014)]{Zhao2014}
Wang, Q.M., Zhao, X.H., 2014. 
Phase diagrams of instabilities in compressed film-substrate systems. 
J. Appl. Mech. 81(5), 051004.

\bibitem[Wu et al.(2017)]{Wu2017}
Wu, B., Su, Y.P., Chen, W.Q., Zhang, C.Z., 2017. 
On guided circumferential waves in soft electroactive tubes under radially inhomogeneous biasing fields. 
J. Mech. Phys. Solids 99, 116-145.

\bibitem[Wu et al.(2013)]{Wu2013}
Wu, Z.L., Moshe, M., Greener, J., Therien-Aubin, H., Nie, Z., Sharon, E., Kumacheva, E., 2013. 
Three-dimensional shape transformations of hydrogel sheets induced by small-scale modulation of internal stresses. 
Nat. Commun. 4, 1586.

\color{black}
\bibitem[Yang et al.(2017)]{Yang2017}
Yang, S.Y., Zhao, X.H., Sharma, P., 2017. 
Avoiding the pull-in instability of a dielectric elastomer film and the potential for increased actuation and energy harvesting. 
Soft Matter 13(26), 4552-4558.
\color{black}

\bibitem[Yin et al.(2018)]{Yin2018}
Yin, S.F., Li, B., Cao, Y.P., Feng, X.Q., 2018. 
Surface wrinkling of anisotropic films bonded on a compliant substrate. 
Int. J. Solids Struct. 141, 219-231.

\color{black}
\bibitem[Zhao and Suo(2007)]{Zhao2007}
Zhao, X.H., Suo, Z.G., 2007. 
Method to analyze electromechanical stability of dielectric elastomers. 
Appl. Phys. Lett. 91(6), 061921.
\color{black}


\end{thebibliography}
\end{document}